# Plasma Internal Inductance Dynamics in a Tokamak

J.A. Romero[1] and JET-EFDA Contributors*.

## JET-EFDA, Culham Science Centre, OX14 3DB, Abingdon, UK

[1]*Asociación Euratom-Ciemat para la fusión*, Madrid, Spain.
*See the Appendix of F. Romanelli et al., Proceedings of the 22nd IAEA Fusion Energy Conference 2008, Geneva, Switzerland

### Abstract—

A lumped parameter model for tokamak plasma current and inductance time evolution as function of plasma resistance, non-inductive current drive sources and boundary voltage or poloidal field (PF) coil current drive is presented. The model includes a novel formulation leading to exact equations for internal inductance and plasma current dynamics. Having in mind its application in a tokamak inductive control system, the model is expressed in state space form, the preferred choice for the design of control systems using modern control systems theory. The choice of system states allow many interesting physical quantities such as plasma current, inductance, magnetic energy, resistive and inductive fluxes etc be made available as output equations.

The model is derived from energy conservation theorem, and flux balance theorems, together with a first order approximation for flux diffusion dynamics. The validity of this approximation has been checked using experimental data from JET showing an excellent agreement.

## I. INTRODUCTION

Tokamaks are pulsed devices modelled as a toroidal transformer with one turn secondary R, L plasma ring circuit coupled with a primary transformer circuit, where R,L denote plasma resistance and inductance respectively. The coupling between transformer primary and plasma is accounted by a mutual inductance M.

The inductance of a conventional electrical system is a parameter depending solely on geometrical factors, but at high frequencies non geometrical effects arise as a result of the slow flux penetration inside the conductor, or skin effect [1],[2]. These are taken into account by decomposing the inductance into a geometry dependent part, the external inductance, and a frequency dependent part, the internal inductance. External and internal contributions also account for energy stored in the magnetic field outside and inside the conductor.

Due to the small size of conventional circuits conductors, the skin effect in conventional circuits starts to be taken into consideration at relatively high frequencies.

The time for flux penetration in tokamaks, however, ranges from a fraction of a second to several seconds, due to the large machine size (several meters) and high temperature (several keV). A further difference is that inductance in conventional circuits is analysed in the context of AC excitation and frequency response, while tokamaks are operated in just half a cycle, and state space time domain model is more appropriate for the analysis.

In a tokamak, an internal inductance accounts for the energy stored in the poloidal field created by the plasma current and external poloidal field currents, while a mutual inductance accounts for the flux linkage between primary inductive coils and the secondary, which is the plasma ring itself. An equivalent ohmic resistance accounts for the Joule losses in the plasma [3],[4] .The internal inductance in a tokamak evolves as the magnetic flux and associated internal current density distribution diffuses in the plasma, and also as the external equilibrium field imposed by external poloidal field coils evolves to maintain the plasma within the vacuum vessel boundaries.

There has been a recent interest in the tokamak community to predict and/or control the plasma inductance behaviour. Control of internal inductance at a low value is required to extend the duration of tokamak plasma discharges with a limited amount of flux in the transformer primary circuit [5] , [6], to reduce the growth rate of the vertical instability of elongated plasmas [7] , [8]  , [9] and to guarantee access to advanced tokamak scenarios with limited amount of flux available at the transformer primary circuit [10] . Tokamak Inductive control has also been shown to be able to shape q profiles and maintain internal transport barriers [11] .

The development of these control systems starts by obtaining a lumped parameter models that approximate processes best described by distributed parameter simulations [12] . To be able to use modern control theory, these models must describe the time evolution of the controlled variables as function of the available actuator and disturbance inputs using the state space formalism.

The basic lumped parameter model describing the tokamak as a toroidal transformer was established from the early days of tokamak fusion research. It has been used to design plasma current control systems using only the expected or nominal values of R,L,M ensuring by design the control system's performance and stability in the face small variations of these parameters due to changing plasma temperature and geometry. From the control point of view, the small variations on the plasma inductance  were treated as perturbations inside the feedback loop. So far, this has been a practical approach, since lumped parameters models for the plasma inductance dynamics have not existed despite more than 50 years of tokamak research.
Full tokamak inductive control, however, requires modelling of both plasma current and inductance. For this, lumped parameter models describing inductance dynamics are required in addition.
Previous modelling the internal inductance dynamics has been restricted to distributed parameter simulations. These are useful for predictions and scenario development, but as it was stated earlier, they are not very useful to design control systems. Modern control theory requires state space models or transfer functions to design the controls.

Having in mind its application in a tokamak inductive control system,  this work develops such a lumped parameter state space model describing self consistently the dynamics of plasma current and inductance as function of plasma resistance, non-inductive current drive sources and boundary loop voltage / PF coil current time derivatives.

The model is expressed in state space form, the preferred choice for the design of control systems using modern control systems theory. The choice of system states allow many interesting physical quantities such as plasma current, inductance, magnetic energy, resistive and inductive fluxes etc be made available as output equations. The validity of this model has been checked using experimental data from JET showing an excellent agreement.

The usefulness of the lumped parameter model presented is demonstrated by developing a mathematical expression for the existing correlation between plasma current ramp rates and internal inductance changes. It will be shown that, contrary to what is commonly believed, plasma current ramp rates are not the cause of internal inductance changes.

The state space model presented can be readily used to design advanced tokamak control systems. It can also be used to dimension tokamak primary transformer circuits from very simple estimations of expected plasma parameters. The models are also useful for other fields of plasma research where toroidal configurations are used, such as Inductively Coupled Plasma sources.

The paper is organised as follows. Section II outlines the essentials of the tokamak transformer. The mathematical derivation used leads to a transformer equation in which the interpretation of plasma inductance as the sum of internal plus external components becomes clear. Some useful intermediate results are obtained through a derivation of Poynting´s theorem in tokamak geometry that differs from the conventional approach found in text books.
Section III uses the results obtained in the transformer modelling to derive exact internal inductance and plasma current state space equations. A first order approximation for flux diffusion dynamics is also presented.

The mathematical derivation of the transformer model and the state space equations is cumbersome. The final result, however, is quite simple. We invite the reader to jump to the internal inductance and plasma current state space equations (39), (40) to appreciate its simplicity.

Section IV outlines an alternative version of the state space model in which a flux diffusion approximation formulated in integral form. This allow us to write a compact state space model describing the plasma current and internal inductance dynamics as function of the external coil drive, found in section V. This alternative formulation is also used later in the paper to find the exact correlation between plasma inductance and plasma current ramp rates.
In section VI the state space model is validated against JET discharges.
In section VII, a common misunderstanding that leads to believe that plasma current ramp rates are the cause of internal inductance changes will be examined, followed by the main conclusions. Some mathematical derivations used in the model construction are included in later appendixes.

## II.   THE TOKAMAK AS A TRANSFORMER

This section outlines the standard Poynting´s and flux balance analysis applied to a tokamak [3] . A cylindrical coordinate system is used $(r, \phi, z)$ and the plasma is assumed to be axis-symmetric around the z-axis. Only the time evolving components $\left(B_r, B_z\right)$ of the *poloidal* magnetic field are considered in the analysis.

The region of integration will be delimited by the region where there is a plasma. This will correspond to a plasma volume $G$, or a plasma cross section $\Omega$, or a plasma boundary $\Gamma$.

The magnetic energy stored in the plasma volume $G$ is obtained from poloidal magnetic field as

$$W = \frac{1}{2\mu_0} \int_G \left(B_r^2 + B_z^2\right) dv \qquad (1)$$

Where $\mu_0$ is the vacuum magnetic permeability and a differential volume element

$$dv = r\, dr\, d\phi\, dz \qquad (2)$$

This contains magnetic field created by the plasma current as well as magnetic field created by external conductors.

Using the vector potential with Coulomb gauge

$$\mathbf{A} = \left(A_r \quad A_\phi \quad A_z\right) = \left(0 \quad \frac{\psi}{2\pi r} \quad 0\right) \qquad (3)$$

where $\psi$ is the flux through an arbitrary circle of radius r centred at the torus symmetry axis, the magnetic field can be obtained from a vector potential, as

$$\mathbf{B} = \nabla \times \mathbf{A} \qquad (4)$$

This renders the usual expressions for magnetic field components in a tokamak

$$B_r = -\frac{\partial A_\phi}{\partial z} = -\frac{1}{2\pi r}\frac{\partial \psi}{\partial z}$$

$$B_z = \frac{1}{r}\frac{\partial \left(rA_\phi\right)}{\partial r} = \frac{1}{2\pi r}\frac{\partial \psi}{\partial r} \qquad (5)$$

Similarly, the toroidal current density is obtained from magnetic field as

$$\mu_0 \mathbf{j} = \nabla \times \mathbf{B} \qquad (6)$$

Using the vector identity

$$B^2 = \nabla \cdot \left(\mathbf{A} \times \mathbf{B}\right) + \mu_0 \mathbf{A} \cdot \mathbf{j} \qquad (7)$$

the magnetic energy (1) can then be written as

$$W = \frac{\int_G \mathbf{A} \cdot \mathbf{j}\, dv - \psi_B I}{2} \qquad (8)$$

or in terms of flux [12]

$$W = \frac{\int_\Omega \psi\, j\, dS - \psi_B I}{2} \qquad (9)$$

where $dS = dr\, dz$, $j$ is the toroidal current density, $\psi_B$ is the flux at the plasma boundary $\Omega$ and $I$ is the total plasma current enclosed by this boundary.

$$I = \int_\Omega j\, dS \qquad (10)$$

Details of the derivation of (9) are found in appendix A.

Using Lenz´s law, the voltage at any location is obtained from flux as

$$V = -\frac{d\psi}{dt} \tag{11}$$

And in particular, the boundary loop voltage is

$$V_B = -\frac{d\psi_B}{dt} \tag{12}$$

Time derivative of (9) leads to the Poynting´s theorem

$$\frac{dW}{dt} + \int_\Omega jV dS = V_B I \tag{13}$$

To obtain (13) we have used the identity

$$\mu_0 \mathbf{A} \cdot \frac{\partial \mathbf{j}}{\partial t} = \mathbf{A} \cdot \left( \nabla \times \frac{\partial \mathbf{B}}{\partial t} \right) = -\nabla \cdot \left( \mathbf{A} \times \frac{\partial \mathbf{B}}{\partial t} \right) + \frac{\partial \mathbf{B}}{\partial t} \cdot \left( \nabla \times \mathbf{A} \right) \tag{14}$$

and an integration over the plasma volume to obtain

$$\int_\Omega \psi \frac{\partial j}{\partial t} dS = \int_G A \frac{\partial j}{\partial t} dv = \psi_b \frac{dI}{dt} + \frac{dW}{dt} \tag{15}$$

Details of the derivation of (13), (15) are found in the appendix B.
The toroidal current density can be written in terms of ohmic and non inductive current drive components. Define $\eta$ and $\hat{j}$ as effective plasma resistivity and non-inductive current density in the toroidal direction. Then, ohms law is written as

$$E = \eta \left( j - \hat{j} \right) \tag{16}$$

Define plasma resistance as

$$R = \frac{\int_\Omega \eta j^2 dS}{I^2} \tag{17}$$

Define non inductive current drive fraction as

$$\frac{\hat{I}}{I} = \frac{\int_\Omega j\eta \hat{j} dS}{\int_\Omega \eta j^2 dS} \tag{18}$$

And define an ideal non inductive voltage source equivalent

$$\hat{V} = R\hat{I} = \frac{\int_\Omega j\eta \hat{j} dS}{I} \tag{19}$$

These definitions lead to the circuit equation

$$\frac{dW}{dt} + V_R I = V_B I \tag{20}$$

The resistive voltage drop can be written in terms of the total plasma current and an equivalent non-inductive current $\hat{I}$ or voltage $\hat{V}$ equivalent source.

$$V_R = R\left( I - \hat{I} \right) = RI - \hat{V} \tag{21}$$

Following electrical engineering standards, the internal inductance is defined from the magnetic energy W stored in the poloidal field in the region enclosed by the plasma boundary

$$L_i = \frac{2W}{I^2} \tag{22}$$

Leading to

$$V_B = V_R + \frac{1}{I}\frac{d}{dt}\left(\frac{1}{2}L_i I^2\right) \tag{23}$$

The inductive voltage is defined as

$$V_{ind} = V_B - V_R = \frac{1}{I}\frac{d}{dt}\left(\frac{1}{2}L_i I^2\right) \tag{24}$$

Time integration of (23) leads to

$$\psi_B = \psi_R + \psi_{ind} \tag{25}$$

Where the inductive and resistive fluxes in (25) are identified from (23) as [3]

$$\psi_R = -\int_0^t V_R dt = -\int_0^t R\left(I - \hat{I}\right) dt \tag{26}$$

$$\psi_{ind} = -\int_0^t V_{ind} dt = -\int_0^t \frac{1}{I}\frac{d}{dt}\left(\frac{1}{2}L_i I^2\right) dt = -L_i I + \frac{1}{2}\int_0^t I\frac{dL_i}{dt} dt \tag{27}$$

The sign criteria in the above equations differs from the one given in [3]. In our case, is given by Lenz's law (11) and Ohm's law (21) written in cylindrical coordinates. With this sign convention a boundary flux that increases in time will generate a negative boundary loop voltage and a negative plasma current. The applied boundary flux is invested according to (25) in inductive (27) and resistive (26) flux components.

Finally, the flux at the plasma boundary can be written as the sum of the flux due to the plasma internal current density and the flux due to the external PF system [13]

$$\psi_b = L_e I + \sum M_j I_j \tag{28}$$

where $L_e$ is the plasma external inductance and $M_j$ are the mutual inductances between PF coils and plasma.

The mutual inductance $M_j$ is function of the coil and plasma boundary geometry. It is defined from the line integral of the vector potential $A_j$ due to the coil system j along a field line covering the plasma boundary:

$$M_j = \frac{\int_\Omega A_j dl}{I_j N} \tag{29}$$

Where N the number of turns of the field line around the machine symmetry axis.

The external inductance is similarly defined from the line integral of the vector potential $A$ due to the plasma current distribution along a field line covering the plasma boundary.

$$L_e = \frac{\int_\Omega A dl}{I N} \tag{30}$$

The external inductance is mainly a function of the plasma boundary geometry, with a weak dependence on the flux gradient at the plasma boundary [14].

Combining (25), (28), (27) we obtain

$$\left(L_e + L_i\right) I + \sum M_j I_j = \psi_R + \frac{1}{2}\int_0^t I\frac{dL_i}{dt} dt \tag{31}$$

Which is a transformer equation in which the plasma secondary has an equivalent plasma inductance

$$L_p = L_e + L_i \tag{32}$$

This transformer equation is more easily recognised if we fix constant the plasma inductance and take time derivatives

$$-\sum_j M_j \frac{dI_j}{dt} = R\left(I - \hat{I}\right) + L_P \frac{dI}{dt} \tag{33}$$

The change of flux produced by external coils generates a voltage that compensates the resistive drop and builds up the plasma current.

## III.  STATE SPACE MODEL

We are after a state space description of the plasma with current $I$ and internal inductance $L_i$ as output variables and plasma resistance, non inductive current drive, and boundary loop voltage as inputs.

We start by introducing the current density weighted flux average

$$\psi_C = \frac{\int_\Omega \psi j \, dS}{I} \tag{34}$$

This flux depends on the particular plasma flux and current profile shapes, or equilibrium. We will refer to the *equilibrium flux surface* as the flux surface corresponding to $\psi_C$.

Using (9), (34) the poloidal field magnetic energy can then be written as

$$W = \frac{(\psi_C - \psi_B) I}{2} \tag{35}$$

And using (22)

$$L_i I = (\psi_C - \psi_B) \tag{36}$$

This implies $\psi_C < \psi_B$ for negative plasma current.

Time derivative of (34) leads to

$$\psi_C \frac{dI}{dt} - \int_\Omega \psi \frac{\partial j}{\partial t} dS = (V_C - V_R) I \tag{37}$$

Where the voltage at the equilibrium flux surface is

$$V_C = -\frac{d\psi_C}{dt} \tag{38}$$

And using (15) , (20)  and (23) we finally obtain

$$I \frac{dL_i}{dt} = 2(V_R - V_C) \tag{39}$$

$$L_i \frac{dI}{dt} = V_B + V_C - 2V_R \tag{40}$$

These are exact equations not found in previous literature. They govern plasma current and internal  inductance dynamics as function of the applied boundary voltage, plasma resistive voltage and voltage at the equilibrium flux surface.

Internal inductance reaches steady state conditions when $V_C = V_R$. The steady state solution for the full set of equations corresponds with $V_C = V_R = V_B$, or a constant loop voltage profile across the plasma.

To complete the model we must find a third equation for the equilibrium voltage $V_C$ as function of the applied boundary voltage and resistive drop changes. Flux diffusion evolves to achieve a constant loop voltage profile that equals the boundary loop voltage. A first order approximation for this process is obtained by writing

$$\frac{d(V_C - V_B)}{dt} \cong -\frac{(V_C - V_B)}{\tau} + \frac{k}{\tau}(V_R - V_B) \tag{41}$$

Where $k, \tau$ are a gain and a time constant. The validity of (41) will be checked in a later section. Regardless of the approximation used, the inductance evolves as result of the competition between resistive drop voltage and voltage at the equilibrium flux surface, according to (39).

The approximation (41) can be incorporated in the state space model by making the change of variables

$$V = V_C - V_B \tag{42}$$

$$I\frac{dL_i}{dt} = 2(V_R - V_B) - 2V \tag{43}$$

$$L_i\frac{dI}{dt} = 2(V_B - V_R) + V \tag{44}$$

$$\frac{dV}{dt} \cong -\frac{V}{\tau} + \frac{k}{\tau}(V_R - V_B) \tag{45}$$

The equations (43),(44), (45) define a $3^{\text{th}}$ order state space system as function of the inductive voltage $V_B - V_R$. The model parameters $\{k, \tau\}$ can be found by running an optimization algorithm that search in the parameter space to find the best match to experimental data. This will be shown in a later section.

## IV.  ALTERNATIVE STATE SPACE MODEL FORMULATION

The model can be written in an alternative form if we integrate (41) from an initial time $t = t_0$

$$V_C - V_B \cong \frac{(\psi_C - \psi_B)}{\tau} - \frac{k}{\tau}(\psi_R - \psi_B) + C \tag{46}$$

Where the integration constant C is

$$C = \Big((\psi_C(t_0) - \psi_B(t_0)) + k(\psi_R(t_0) - \psi_B(t_0))\Big)\Big/\tau - (V_C(t_0) - V_B(t_0)) \tag{47}$$

Relative to the initial conditions we can write

$$V_C - V_B \cong \frac{(\psi_R - \psi_B)}{\tau}\left(\frac{(\psi_C - \psi_B)}{(\psi_R - \psi_B)} - k\right) \tag{48}$$

Which is just the integral formulation of the derivative approximation (41).
To obtain the model in compact form, we introduce the state space vector

$$x = (x_1, x_2, x_3)^T \tag{49}$$

with

$$x_3 = \frac{(\psi_C - \psi_B)}{(\psi_R - \psi_B)} \tag{50}$$

$$x_2 = x_3 I \tag{51}$$

$$x_1 = \frac{L_i}{x_3^2} \tag{52}$$

With these new variables the magnetic energy is

$$W_m = \frac{L_i I^2}{2} = \frac{x_1 x_2^2}{2} \tag{53}$$

And the inductive flux is

$$\psi_{ind} = -(\psi_R - \psi_B) = -x_1 x_2 \tag{54}$$

Differentiation of the states using (39), (40), (11) and recursively writing the result as function of the states leads after some algebra to the following state equations

$$\frac{dx_1}{dt} = \frac{2(x_3-1)}{x_2 x_3}\left(V_b - \frac{Rx_2}{x_3} + R\hat{I}\right) \tag{55}$$

$$\frac{dx_2}{dt} = \frac{(2-x_3)}{x_1 x_3}\left(V_b - \frac{Rx_2}{x_3} + R\hat{I}\right) \tag{56}$$

$$\frac{dx_3}{dt} \cong \frac{(k-x_3)}{\tau} - \frac{x_3}{x_1 x_2}\left(V_b - \frac{Rx_2}{x_3} + R\hat{I}\right) \tag{57}$$

The first two equations (55), (56) are exact. The last equation (57) is obtained by writing the approximation (48) as function of the state variables

$$V_C = \frac{(x_3-k)x_1 x_2}{\tau} + V_B \tag{58}$$

and then substituting the result in the exact differential equation for the state $x_3$.

$$\frac{dx_3}{dt} = \frac{(1-x_3)}{x_1 x_2}\left(V_b - \frac{Rx_2}{x_3} + R\hat{I}\right) + \frac{1}{x_1 x_2}\left(\frac{Rx_2}{x_3} - R\hat{I} - V_c\right) \tag{59}$$

The system of equations (55), (56) and (57) responds to an inductive voltage input encompassing external boundary voltage stimuli, plasma resistance changes and non inductive current drive sources.

$$u_1\left(V_b, R, \hat{I}, x\right) = V_b - \frac{Rx_2}{x_3} + R\hat{I} \tag{60}$$

The state space equations (55), (56) and (57) can be integrated in time starting from some given initial conditions for the states, and together with the output equations

$$y = \left(y_1, y_2\right)^T = \left(L_i, I\right)^T \tag{61}$$

$$L_i = x_1 x_3^2 \tag{62}$$

$$I = \frac{x_2}{x_3} \tag{63}$$

constitute an alternative formulation that is equivalent to the one given by (43),(44), (45) . Both models produce identical results. The difference is that the approximation for flux diffusion is given in differential form in one case, and in integral form in the other.

Following the standards for non-linear systems [22], the model can be written in a more compact form as

$$\frac{dx}{dt} = f_1(x) + g_1(x)u_1\left(V_b, R, \hat{I}, x\right)$$

(64)

$$y = h(x)$$

With $f_1(x) = \left(0, 0, \frac{(k - x_3)}{\tau}\right)^T$

(65)

$$g_1(x) = \left(\frac{2(x_3 - 1)}{x_2 x_3} \quad \frac{(2 - x_3)}{x_1 x_3} \quad -\frac{x_3}{x_1 x_2}\right)^T$$

(66)

$$h(x) = \left(x_1 x_3^2, \frac{x_2}{x_3}\right)$$

(67)

This model has the inductance and plasma current as output variables by choice. Using the state space formalism, any function of the states and inputs can be made available as an output equation. For instance magnetic energy (53), inductive flux (54), voltage at the equilibrium flux surface (58) , etc, can be made available by writing the corresponding functions of the states and inputs as model outputs.

Also, augmenting the model with new states such as

$$\frac{dx_4}{dt} = -V_b$$

(68)

The boundary and resistive fluxes can be also be made available as output equations

$$\psi_B = x_4$$

(69)

$$\psi_R = x_1 x_2 + x_4$$

and from here, the Ejima coefficient [3],[5] can also be obtained as an output equation

$$C_E = \frac{\psi_R}{\mu_0 r_0 I} = \frac{(x_1 x_2 + x_4) x_3}{\mu_0 r_0 x_2}$$

(70)

where $r_0$ is the magnetic axis coordinate.

Also, if the magnetic axis position is known, an output equation for the dimensionless internal inductance [15] could be made available as

$$l_i = \frac{4W}{\mu_0 r_0 I^2} = \frac{2L_i}{\mu_0 r_0} = \frac{2x_1 x_3^2}{\mu_0 r_0}$$

(71)

This last normalization is the standard used for the ITER design [6].

## V.  STATE SPACE MODEL AS FUNCTION OF POLOIDAL FIELD CURRENTS

Finally, we have to write the model as function of the PF coil currents surrounding the plasma.

For constant $M_j$, $L_e$ (fixed plasma geometry), Lenz´s law applied to the boundary flux balance (28) leads to

$$V_B = -L_e \frac{dI}{dt} - M_j \frac{dI_j}{dt}$$

(72)

Which combined with (40) leads to

$$V_B = \left(\frac{-L_i}{(L_i + L_e)}\right) M_j \frac{dI_j}{dt} - \frac{L_e}{(L_i + L_e)}(V_C - 2V_R)$$

(73)

And using (21) (58),(62),(63), the inductive voltage (60) can be written as

$$\left(V_b - \frac{Rx_2}{x_3} + R\hat{I}\right) = -\frac{L_e}{\left(x_1 x_3^2 + 2L_e\right)}\left(\frac{(x_3-k)x_1 x_2}{\tau}\right) + \frac{x_1 x_3^2}{\left(x_1 x_3^2 + 2L_e\right)}\left(-\sum_{j=1}^{N} M_j \frac{dI_j}{dt} - \frac{Rx_2}{x_3} + R\hat{I}\right) \quad (74)$$

The validity of this expression is conditioned to the validity of the approximation (58). The inductive voltage (74) can then be incorporated into the state equations (55), (56) and (57) , and the state space model can then finally be written in compact form as

$$\frac{dx}{dt} = f_2(x) + g_2(x)u_2\left(V_b, R, \hat{I}, x\right) \quad (75)$$

$$y = h(x)$$

With $h(x)$ given by (67) , and

$$u_2\left(I_j, R, \hat{I}, x\right) = -\sum_{j=1}^{N} M_j \frac{dI_j}{dt} - \frac{Rx_2}{x_3} + R\hat{I} \quad (76)$$

$$g_2(x) = \frac{x_1 x_3^2}{\left(x_1 x_3^2 + 2L_e\right)} g_1(x) \quad (77)$$

$$f_2(x) = -\frac{L_e}{\left(x_1 x_3^2 + 2L_e\right)}\left(\frac{(x_3-k)x_1 x_2}{\tau}\right) g_1(x) + f_1(x) \quad (78)$$

Where the new input to the state space equations is now a function of the PF coil drive, plasma resistance and non inductive current drive. Same procedure can be applied to the case where external and mutual inductances are function of time, resulting in an additional input to (76).

$$u_2\left(I_j, R, \hat{I}, x\right) = -\sum_{j=1}^{N} M_j \frac{dI_j}{dt} - \frac{Rx_2}{x_3} + R\hat{I} - \sum_{j=1}^{N} I_j \frac{dM_j}{dt} - I \frac{dL_e}{dt} \quad (79)$$

## VI.  STATE SPACE MODEL VALIDATION

To validate the state space model we use the actual readings of real time diagnostics at the JET tokamak. Some of plasma states and inputs can not be directly measured. The data used for the validation is obtained from a real time magnetic software at JET called FELIX [16] . This code, evolved from the previous code XLOC [17] , provides the magnetic vacuum topology reconstruction using magnetic measurements obtained outside the plasma. Alternatively, we could have used off-line equilibrium codes such as EFIT [18] . The reason to use FELIX instead of EFIT is motivated by the fact that FELIX was foreseen to be used in the control application for which the presented model was developed [19] . Anyhow, FELIX outputs have been tested over the  years to match closely the EFIT outputs at JET.  Among other plasma variables, FELIX delivers plasma current, boundary loop voltage, normalised internal inductance, ohmic power and magnetic axis coordinates.  Plasma internal inductance is derived from magnetic axis position and *normalised* internal inductance, using (71).  Plasma resistance is derived from ohmic power, and the equilibrium fluxes and voltages from  (36) and (38) respectively.

Plasma resistance and boundary voltage are inputs the state space model given by equations (55), (56) , (57), with an initial guess for the initial conditions and the adjustable parameters $\{k, \tau\}$ . The current and inductance outputs of the state space model are then compared with the actual JET data and an optimization algorithm is used to search on the initial conditions and parameter space to find the best match in the Akaike´s final prediction error sense [21] .

The figures below show simulation results for an optimized cases with fixed $k \cong 0.98$ and $\tau \cong 1.25$, for two discharges with step up/down on plasma current in the flat top and negligible current drive. The first order approximation for the flux diffusion process along with the non linear relationships in the state space model are sufficient to reproduce the experimental data with reasonable accuracy. Of course running the optimization using data segmentation for the three distinct phases (ramp-up, flat top and ramp-down) can increase the accuracy of the simulations providing different sets of parameters for each segment. But the interesting point here is that a first order approximation with two parameters $\{k, \tau\}$ can reproduce most of the experimental data with reasonable accuracy.

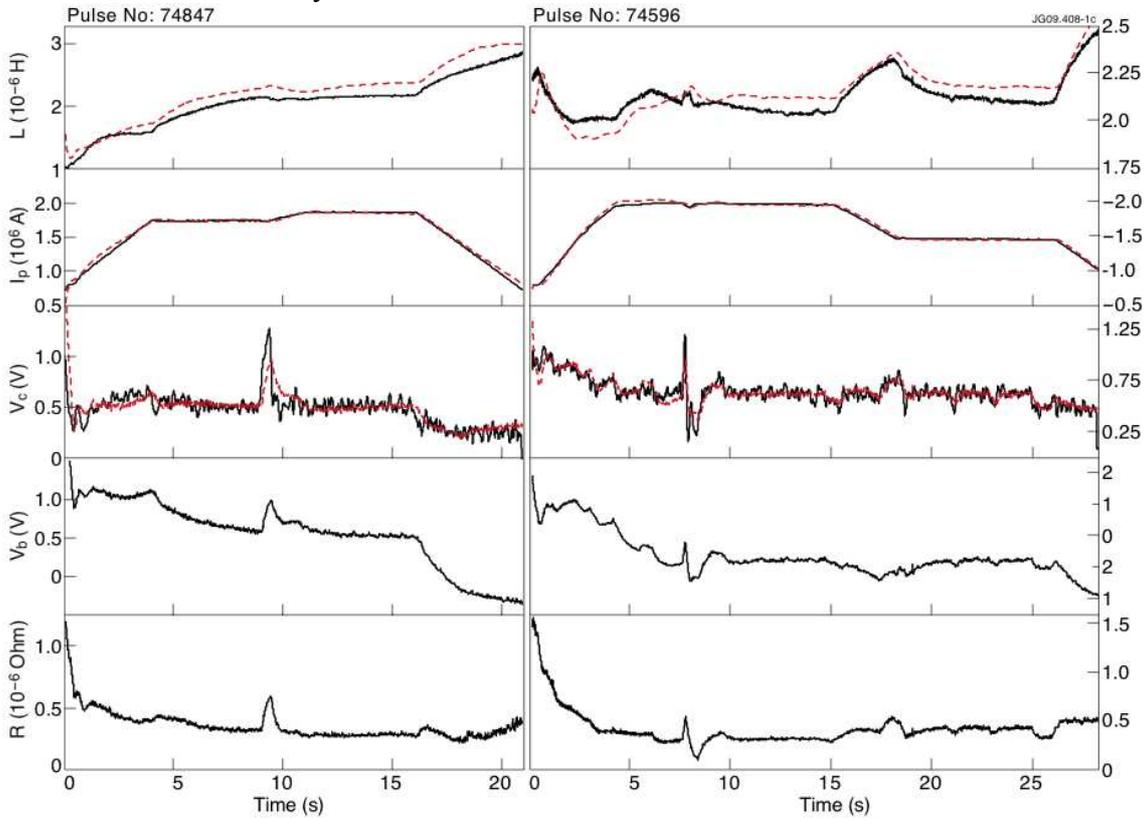

Fig. 1. Comparison between experimental readings (black) and state space model outputs (red). In top-down order are shown the plasma internal inductance, plasma current, voltage $V_C$ at the equilibrium flux surface $\psi_C$, boundary voltage and plasma resistance.

## VII. RELATIONSHIP BETWEEN INDUCTANCE AND PLASMA CURRENT RAMP-RATE

Taking the ratio between (55) and (56)

$$\frac{1}{x_1}\frac{dx_1}{dt} = \frac{2(x_3 - 1)}{(2 - x_3)}\frac{1}{x_2}\frac{dx_2}{dt} \qquad (80)$$

And using (52),(51) and (50) we arrive to

$$\frac{dL_i}{dt} = \frac{2(x_3 - 1)}{(2 - x_3)}\frac{L_i}{I}\frac{dI}{dt} + \frac{2L_i}{x_3(2 - x_3)}\frac{dx_3}{dt} \qquad (81)$$

Taking time derivative of (50), steady state conditions for $x_3$ are obtained when

$$(V_R - V_B) x_3^0 = (V_C - V_B) \tag{82}$$

Where $x_3^0$ is the steady state value. According to (39), steady state conditions for inductance are obtained when $V_C = V_R$. While steady state condition for $L_i$ implies steady state condition for $x_3$, the converse is not necessary true. It is possible to obtain steady state evolution for $x_3$ and not for $L_i$. Only in this situation an strict correlation between plasma current ramp rates and internal inductance changes exists.

$$\frac{dL_i}{dt} = \frac{2(x_3^0 - 1)}{(2 - x_3^0)} \frac{L_i}{I} \frac{dI}{dt} \tag{83}$$

A common misunderstanding or language abuse is to state that the internal inductance changes are produced by plasma current ramp rates. Equations (81) or (83) quantify a correlation between plasma current ramp rates and internal inductance changes, but it does not imply a cause effect relationship between current ramp rates and inductance changes. Both plasma current (40) and inductance (39) have a common cause, which is the applied inductive voltage (24) to the plasma . Because they have a common cause, they exhibit a correlation. But the internal inductance evolves depending on the competition between resistive drop voltage (21) and voltage (38) at the equilibrium flux surface, according to (39). The existing correlation, however, has successfully been exploited to control the internal inductance using the plasma current ramp rate as a virtual actuator [6]. In this case, an internal inductance error signal becomes a reference for plasma current by means of a time integration of an empirical version of (83) . The reference for the plasma current is sent to the plasma current control system that uses the primary of the transformer as the actuator. A more direct option is to use directly the transformer coil as the actuator. In any case, the state space models presented can be used as the keystone for the design of advanced non linear controllers [19] , [22].

## VIII. Conclusions

Using a first order approximation for flux diffusion dynamics together with energy conservation and flux balance theorems, a non linear model for plasma current and inductance time evolution as function of plasma resistance, non-inductive current drive and boundary loop voltage / PF coil current time derivatives has been obtained. The model is expressed in state space form, the preferred choice for the design of control systems using modern control systems theory. The choice of system states allow many interesting physical quantities such as plasma current, inductance, magnetic energy, resistive and inductive fluxes etc be made available as output equations. The validity of this model has been checked using experimental data from JET showing an excellent agreement.
Contrary to what is commonly believed, plasma current ramp rates are not the cause of internal inductance changes, although both are strongly correlated under some circumstances. A mathematical expression for this correlation has been derived from the state space model.

## IX. Acknowledgment / Disclaimer

The authors are very grateful to UPV/EHU and the Science and Innovation Council MICINN for its support through research projects GIU07/08, ENE2009-07200 and ENE2010-18345 respectively They are also grateful to the Basque Government for its partial support through the research projects S-PE07UN04, S-PE08UN15 and S-PE09UN14.

This work, supported by the European Communities under the contract of Association between EURATOM and Ciemat, was carried out within the framework of the European Fusion Development Agreement. The views and opinions expressed herein do not necessarily reflect those of the European Commission.

## APPENDIX A

It follows the derivation of equation (9).

The magnetic energy stored in the plasma region is

$$W = \frac{1}{2\mu_0} \int_G B^2 \, dv \tag{84}$$

Using the vector identity

$$B^2 = \nabla \cdot (\mathbf{A} \times \mathbf{B}) + \mu_0 \mathbf{A} \cdot \mathbf{j} \tag{85}$$

The magnetic energy can be written as

$$W = \frac{1}{2\mu_0} \int_G \nabla \cdot (\mathbf{A} \times \mathbf{B}) \, dv + \frac{1}{2} \int_G \mathbf{A} \cdot \mathbf{j} \, dv \tag{86}$$

Gauss theorem applied to first term on the right hand side of (86) leads to

$$\frac{1}{2\mu_0} \int_G \nabla \cdot (\mathbf{A} \times \mathbf{B}) \, dv = \frac{1}{2\mu_0} \int_\Gamma (\mathbf{A} \times \mathbf{B}) \cdot \mathbf{dS} \tag{87}$$

For the poloidal components of the **B** field and toroidal component of vector potential **A**, the vector product is reduced to

$$(\mathbf{A} \times \mathbf{B}) = \frac{\psi}{2\pi r} \begin{pmatrix} B_Z & 0 & -B_R \end{pmatrix} \tag{88}$$

The differential surface element is parallel to the product (88), and its magnitude is

$$dS = 2\pi r \, dl \tag{89}$$

Where $dl$ is a differential path element.

Then, the surface integral can be transformed into a line integral, and using Ampere's law the surface integration is reduced to

$$\frac{1}{\mu_0} \int_\Gamma (\mathbf{A} \times \mathbf{B}) \cdot \mathbf{dS} = -\psi_B I \tag{90}$$

Combining (86) and (90) we obtain

$$W = \frac{1}{2} \int_G \mathbf{A} \cdot \mathbf{j} \, dv - \frac{\psi_B I}{2} \tag{91}$$

Or exploiting the relationship between flux and vector potential (3) we finally obtain equation (9), which we reproduce again as a courtesy to the reader.

$$W = \frac{\int_{\Omega} \psi \, j \, dS - \psi_B I}{2} \tag{92}$$

<p align="center">APPENDIX B</p>

It follows the derivation of derivation of equations (13), (15).
Time derivative of (9) leads to

$$2\frac{dW}{dt} = \frac{d}{dt}\int_{\Omega} \psi \, j \, dS + V_B I - \psi_B \frac{dI}{dt} \tag{93}$$

Or in terms of vector potential

$$2\frac{dW}{dt} = \frac{d}{dt}\int_{G} \mathbf{A} \cdot \mathbf{j} \, dv + V_B I - \psi_B \frac{dI}{dt} \tag{94}$$

The first term in the right hand side of (94) can be expanded as

$$\frac{d}{dt}\int_{G} \mathbf{A} \cdot \mathbf{j} \, dv = \int_{G} \frac{\partial \mathbf{A}}{\partial t} \cdot \mathbf{j} \, dv + \int_{G} \mathbf{A} \cdot \frac{\partial \mathbf{j}}{\partial t} \, dv \tag{95}$$

Where the first term on the right hand side is the ohmic power input

$$\int_{G} \frac{\partial \mathbf{A}}{\partial t} \cdot \mathbf{j} \, dv = \int_{G} \mathbf{E} \cdot \mathbf{j} \, dv = -\int_{\Omega} j V \, dS \tag{96}$$

The second term in the right hand side of (95) can be expanded as

$$\int_{G} \mathbf{A} \cdot \frac{\partial \mathbf{j}}{\partial t} \, dv = \frac{1}{\mu_0} \int_{G} \mathbf{A} \cdot \left( \nabla \times \frac{\partial \mathbf{B}}{\partial t} \right) dv \tag{97}$$

Using the vector identity

$$\nabla \cdot \left( \mathbf{A} \times \frac{\partial \mathbf{B}}{\partial t} \right) = \frac{\partial \mathbf{B}}{\partial t} \cdot (\nabla \times \mathbf{A}) - \mathbf{A} \cdot \left( \nabla \times \frac{\partial \mathbf{B}}{\partial t} \right) \tag{98}$$

The right hand side of (97) can then be written as

$$\frac{1}{\mu_0} \int_{G} \mathbf{A} \cdot \left( \nabla \times \frac{\partial \mathbf{B}}{\partial t} \right) dv = \frac{1}{\mu_0} \int_{G} \frac{\partial \mathbf{B}}{\partial t} \cdot (\nabla \times \mathbf{A}) dv - \frac{1}{\mu_0} \int_{G} \nabla \cdot \left( \mathbf{A} \times \frac{\partial \mathbf{B}}{\partial t} \right) dv \tag{99}$$

Gauss theorem applied to second term on the right hand side of (99) leads to

$$\frac{1}{\mu_0} \int_{G} \nabla \cdot \left( \mathbf{A} \times \frac{\partial \mathbf{B}}{\partial t} \right) dv = \frac{1}{\mu_0} \oint_{\Gamma} \left( \mathbf{A} \times \frac{\partial \mathbf{B}}{\partial t} \right) \cdot \mathbf{dS} \tag{100}$$

And for the poloidal components of the $\mathbf{B}$ field and toroidal component of vector potential $\mathbf{A}$, the vector product is reduced to

$$\left( \mathbf{A} \times \frac{\partial \mathbf{B}}{\partial t} \right) = \frac{\psi}{2\pi r} \frac{\partial}{\partial t} \begin{pmatrix} B_Z & 0 & -B_R \end{pmatrix} \tag{101}$$

The differential surface element is parallel to the product (101), and its magnitude is given by (89). Then, the surface integral can be transformed into a line integral, and using Ampere's law the surface integration is reduced to

$$\frac{1}{\mu_0} \oint_{\Gamma} \left( \mathbf{A} \times \frac{\partial \mathbf{B}}{\partial t} \right) \cdot \mathbf{dS} = -\psi_B \frac{dI}{dt} \tag{102}$$

Note also that

$$2\mu_0 \frac{dW}{dt} = \frac{d}{dt}\int_{G} B^2 \, dv = 2\int_{G} \mathbf{B} \cdot \frac{\partial \mathbf{B}}{\partial t} \, dv = 2\int_{G} (\nabla \times \mathbf{A}) \cdot \frac{\partial \mathbf{B}}{\partial t} \, dv \tag{103}$$

Combining (97), (99), (100) (102), (103)

$$\int_G \mathbf{A} \cdot \frac{\partial \mathbf{j}}{\partial t} dv = \psi_b \frac{dI}{dt} + \frac{dW}{dt}$$
(104)

Which written in terms of flux is just the equation (15) used for the state space model derivation.

Combining (95) (96), (104) we obtain

$$\frac{d}{dt} \int_G \mathbf{A} \cdot \mathbf{j} dv = -\int_\Omega jV dS + \psi_b \frac{dI}{dt} + \frac{dW}{dt}$$
(105)

And this last equation combined with (94) leads to the Poynting's theorem (13).

$$\frac{dW}{dt} + \int_\Omega jV dS = V_B I$$
(106)